\documentclass[mamp]{jpconf}
\usepackage{amssymb}
\usepackage{epsfig}
\usepackage{bm}
\def\ltap{\ \raisebox{-.4ex}{\rlap{$\sim$}} \raisebox{.4ex}{$<$}\ }

\begin{document}

\title{Spacetime foam at a TeV}

\author{Luis A. Anchordoqui}
\address{Department of Physics,
University of Wisconsin-Milwaukee,\\ 
P.O. Box 413, Milwaukee, WI 53201}

\ead{anchordo@uwm.edu}

\date{August 2006}

\begin{abstract}
\noindent Motivated by recent interest in TeV-scale
gravity and especially by the possibility of fast
baryon decay mediated by virtual black holes, we study
another dangerous aspect of spacetime foam interactions: lepton flavor
violation. We correlate existing limits on gravity-induced
decoherence in the neutrino sector with a lower bound on the scale of
quantum gravity, and find that if spacetime foam interactions do not allow an
$S$-matrix description the UV cutoff is well beyond the
electroweak scale. This suggests that string theory provides the appropriate
framework for description of quantum gravity at the TeV-scale.
\end{abstract}

What is the meaning of quantum gravity? It means that spacetime itself
is subject to quantum laws, necessitating inherent fluctuations in the
fabric (metric and topology) of space and time. These microscopic
boiling bubbles force on spacetime a foam-like
structure~\cite{Wheeler:1957mu}. A heuristic example pictures
spacetime to be filled with tiny virtual black holes that pop in and
out of existence on a timescale allowed by Heisenberg's uncertainty
principle~\cite{Hawking:1995ag}. These black holes conserve energy,
angular momentum, and electric and color charge (unbroken gauged
quantum numbers), but they {\it are believed} not to conserve global
quantum numbers. If this is the case, the transition between initial
and final density matrices associated with black hole formation and
evaporation is not factorizable into products of $S$-matrix elements
and their hermitian conjugates. The evolution of such a quantum system
is characterized by a superscattering operator $\mathbb{S}$ that maps
initial pure states to final mixed states, $\rho_{\rm out} =
\mathbb{S} \rho_{\rm in},$ with $\mathbb{S} \neq S^\dagger
S$~\cite{Hawking:1976ra}. In other words, there {\em may be} a loss of
quantum information across the black hole event horizons, providing an
environment that can induce decoherence of apparently isolated matter
systems.

In recent years much attention has been devoted to TeV-scale gravity
models~\cite{Arkani-Hamed:1998rs}, as they provide an economic
explanation of the hierarchy between the Planck and electroweak mass
scales. In the canonical example, the Standard Model (SM) fields are
confined to a four dimensional world (corresponding to our apparent
universe), while gravity spills into large spatial compact 
dimensions without conflicting with experimental
bounds~\cite{Cullen:1999hc}. Of particular interest is the question
whether fast baryon decay can proceed via virtual black hole states in
the spacetime foam~\cite{Adams:2000za}. The process is envisioned as
the simultaneous absorption of two quarks into the black hole,
followed without memory of the initial state by the thermal emission
of an antiquark and a lepton, entailing a change in the global baryon
and lepton quantum numbers $qq \to \overline q l$. The probability
that two quarks in a proton of size $\Lambda_{\rm QCD}^{-1}$ pass
within a fundamental Planck length, within the Heisenberg lifetime
uncertainty of the black hole is $\propto (\Lambda_{\rm QCD}/M_{\rm
  QG})^4,$ where $M_{\rm QG}$ is the gravitational UV cutoff. Thus,
the present limit on the proton lifetime, $\tau_p \sim
10^{33}$~yr~\cite{Shiozawa:1998si}, implies~\cite{Adams:2000za}
\begin{equation}
 M_{\rm QG} > 10^{16}~{\rm GeV} \,\,.
\end{equation}

Interactions through the higher dimensional QQQL operator can be
prevented if one separates the quark and lepton fields far enough in
an extra dimension so that their wave function overlap is
exponentially suppressed~\cite{Arkani-Hamed:1999dc}. Within this
picture one may imagine that quark and leptons are fields localized on
different branes at the boundaries of a thick wall, but of course
gauge and higgs fields are free to propagate inside the wall so that
quarks and leptons interact with one another through SM interactions.
This division will certainly inhibit fast baryon decay. However,
violation of global quantum numbers for processes with spacetime foam
black holes on the lepton brane should proceed without wave function
suppression. If there is loss of quantum information across event
horizons, an interesting quantitative measurement of interaction with
the spacetime foam emerges in its effect on neutrino oscillations:
interaction with the virtual black holes introduces a decohering
process which can be strongly constrained by current
observations~\cite{Lisi:2000zt}. In this work we correlate the
existing limits on gravity-induced decoherence with a lower bound on
the scale of quantum gravity. We then examine how this bound can shed
light on the nature of TeV-scale gravity.

The SM is based on the gauge group $ G_{\rm SM} = SU(3)_{\rm C} \times
SU(2)_{\rm L} \times U(1)_{\rm Y},$ with three fermion generations. A
single generation consists of five different representations of the
gauge group: $Q_{\rm L}(3,2,\phantom{-}1/6),$ $U_{\rm R}(3,1,2/3),$
$D_{\rm R}(3,1,-1/3),$ $L_{\rm L}(1,2,-1/2),$ $E_{\rm R} (1,1,-1);$
where the numbers in parenthesis represent the corresponding charges
under $G_{\rm SM}.$ The model contains a single higgs boson doublet,
$\phi(1/2,1/2),$ whose vacuum expectation value breaks the gauge
symmetry $G_{\rm SM}$ into $SU(3)_{\rm C} \times U(1)_{\rm EM}.$ An
important feature of the SM, which is relevant to spacetime foam
interactions discussed here, is the fact that the SM also comprises an
accidental global symmetry, $ G_{\rm SM}^{\rm global} = U(1)_B \times
U(1)_e \times U(1)_\mu \times U(1)_\tau,$ where $U(1)_B$ is the baryon
number symmetry, and $U(1)_{e,\mu,\tau}$ are three lepton flavor
symmetries, with total lepton number given by $L = L_e + L_\mu +
L_\tau$. It is an accidental symmetry because we do not impose it. It
is a consequence of the gauge symmetries and the low energy particle
content. It is possible (but not necessary), however, that effective
interaction operators induced by the high energy content of the
underlying theory may violate sectors of the global symmetry. This
violation is already present in the absence of gravity, as has been
evidenced by the discovery of neutrino flavor oscillations. If there
are additional contributions of this kind from the gravity sector,
then black hole-mediated interactions could violate $G_{\rm SM}^{\rm
  global}$ in a distinct manner, as mentioned above and reviewed in
what follows.

Measurements of flavor transformations in a neutrino beam can
provide a clean and sensitive probe of interactions with the
spacetime foam. Without interference from the gravitational
sector, oscillations in the neutrino sector provide a pure quantum
phenomenon in which the density matrix has the properties of a
projection operator, Tr~$\rho^2=$ Tr~$\rho=1.$  Because black
holes do conserve energy, angular momentum (helicity), color and
electric charge, any neutrino interacting with the virtual black
holes needs to re-emerge as a neutrino. As an example, if
spacetime foam black holes do not conserved $U(1)_e \times
U(1)_\mu \times U(1)_\tau,$ neutrino flavor is randomized by
interactions with the virtual black holes. The result of many
interactions is to equally populate all three possible flavors.

The ensuing discussion will be framed in the context of
$\nu_\mu\leftrightarrow\nu_\tau$ oscillations, and we will comment
on other channels after presenting our results.
Consider two neutrino states $\nu_1=(1,0)^T$ and $\nu_2=(0,1)^T$ with masses
$m_1$ and $m_2$, and two  flavor states
$\nu_\mu=(c_\theta,s_\theta)^T$ and
$\nu_\tau=(-s_\theta,c_\theta)^T$, where $\theta$ is the neutrino
mixing angle, $c=\cos$, $s=\sin$, and $T$ denotes the transpose.
In the absence of spacetime foam interactions the time evolution
equation for the density matrix,
\begin{equation}
\dot\rho=-i[H,\rho]\ ,
\label{liouville}
\end{equation}
is governed (in the mass basis) by the Hamiltonian
$H=\textstyle\frac{1}{2}\,{\rm diag}(-k,+k)$,
where  $k=\Delta m^2/2E$,  $\Delta m^2=m^2_2-m^2_1$, and $E\;(\gg m_{1,2})$
is the neutrino energy (in natural units).  The solution $\rho(t)$ of
Eq.~(\ref{liouville}), with initial conditions $\rho(0)=\Pi_{\nu_\mu}$,
gives the $\nu_\mu$ survival probability after propagation of a distance $x$,
\begin{equation}
P(\nu_\mu \to \nu_\mu)  =  {\rm Tr}~[\Pi_{\nu_\mu} \ \rho(t)] 
  = 
1- s^2_{2\theta}\,\,(1-\cos kx)/2\ .
\label{prob}
\end{equation}
Here, $\Pi_{\nu_\mu}=\nu_\mu\otimes\nu_\mu^\dagger$ is the $\nu_\mu$
state projector. Transitions from pure to mixed states become possible
by addition of a dissipative term ${\cal D}[\rho]$ into
Eq.~(\ref{liouville})~\cite{Ellis:1983jz}
\begin{equation}
\dot\rho=-i[H,\rho] - {\cal D}[\rho]\ .
\label{liouvilleD}
\end{equation}
Energy conservation, complete positivity (assuring the absence of
unphysical effects, such as negative probabilities, when dealing
with correlated systems), and monotonic increase in the
von-Neumann entropy eventually lead to a modification of
Eq.~(\ref{prob})~\cite{Benatti:2000ph},
\begin{equation}
P (\nu_\mu \to \nu_\mu) = 1- s^2_{2\theta}\,\, (1-e^{-\gamma x}\cos kx)/2 \ \,.
\label{pdeco}
\end{equation}
Here, $\gamma$ has dimension of energy, and its inverse defines the typical
(coherence) length after which the system gets mixed. Thus,
for $\gamma x \sim {\cal O}(1)$ one expects significant deviations from
Eq.~(\ref{prob}). We
are interested in dissipative scenarios where decoherence
effects vanish in the weak gravitational limit, $M_{\rm
QG}\rightarrow \infty,$ and thus we take
\begin{equation}
\gamma  = \tilde \kappa \,\left(\frac{E}{{\rm GeV}} \right)^n \,
\left(\frac{M_{\rm QG}}{{\rm GeV}}\right)^{-n+1}~{\rm GeV} \,\,,
\label{osde}
\end{equation}
where $\tilde\kappa$ is a dimensionless parameter which by
naturalness is expected to be ${\cal O}(1)$ and $n \geq 2.$

Atmospheric neutrinos are generated in the decay of pions and kaons
resulting from cosmic ray collisions. Most
of them are produced in a spherical surface at about 10-20~km above
ground level and they proceed towards the earth. Production of muon
neutrinos is dominated by the process $\pi^+ \to \mu^+ \nu_\mu$ (and
its charged conjugate). At energies $\ltap 10$~GeV the muons decay
before reaching the surface of the earth, $\mu^+ \to e^+ \overline
\nu_\mu \nu_e.$ This decay chain then leads to a flavor ratio $\nu_e :
\nu_\mu \approx 1: 2.$ The zenith angle distribution observed by
Super-Kamiokande shows a clear deficit of upward-going muon neutrinos,
which is well explained by two flavor $\nu_\mu\leftrightarrow\nu_\tau$
oscillations, with $\Delta m^2 \simeq 3 \times 10^{-3}$~eV$^2$ and
$s^2_{2\theta} \simeq 1$~\cite{Fukuda:1998mi}.

A best fit to data collected by the Super-Kamiokande atmospheric
neutrino experiment~\cite{Fukuda:1998mi}, allowing for both
oscillation and decoherence yields, for $n=2,$
\begin{equation}
  \tilde \kappa \,\, \left(\frac{M_{\rm QG}}{{\rm GeV}}\right)^{-1}
  < 0.9 \times 10^{-27} \,\,, \label{tkappa}
\end{equation}
at the 90\% CL~\cite{Lisi:2000zt}. For larger $n,$ Eq.(\ref{tkappa})
generalizes to~\cite{Anchordoqui:2005is}
\begin{equation}
\tilde \kappa \,\, \left(\frac{M_{\rm QG}}{{\rm
GeV}}\right)^{-n+1}
 < 0.9 \times 10^{-27} \,\, . \label{tkappan}
\end{equation}
Indeed, this represents  a conservative bound. This can be seen
from Eq.~(\ref{osde}): the analysis of data places bounds on
$\gamma,$ and the neutrino energies are well above 1 GeV. From
Eq.~(\ref{tkappan}) it is straightforward to see that, for $\tilde
\kappa \sim 1,$ {\em the lower limit on the UV cutoff is well
beyond the electroweak scale.}

We now comment on other possible oscillation channels. The CCFR
detector at Fermilab is sensitive to $\nu_\mu \to
\nu_e$~\cite{Romosan:1996nh} and $\nu_e \to \nu_\tau$
\cite{Naples:1998va} flavor transitions. Neutrino energies range
from 30 to 600 GeV with a mean of 140 GeV, and their flight
lengths vary from 0.9 to 1.4~km. A best fit to the data allowing
for both oscillation and decoherence yields
\begin{equation}
\tilde \kappa \,\, \left(\frac{M_{\rm QG}}{{\rm GeV}}\right)^{-n+1} 
< 2.0 \times 10^{-24} \,\,,
\end{equation}
at the 99\% CL~\cite{Gago:2000qc}. Moreover, similar constraints on
gravity-induced decoherence arise in the quark sector. The persistence
of coherence in $K^0\overline K^0$ oscillations and neutron
interferometry leads to~\cite{Ellis:1983jz}
\begin{equation}
\tilde \kappa \,\, \left(\frac{M_{\rm QG}}{{\rm GeV}}\right)^{-n+1} 
< 2.0 \times 10^{-21} \,\, .
\end{equation}
The typical energy in these systems is $E \sim 1~{\rm GeV}.$ In the
near future, the possible observation of Galactic anti-neutrino
beams~\cite{Anchordoqui:2003vc} at the IceCube facility may provide a
major improvement in sensitivity to spacetime foam
interactions~\cite{Anchordoqui:2005gj}.

In summary, using an existing analysis of atmospheric neutrino
data we have shown that if TeV-scale gravity is realized in
nature, interactions with virtual black holes are non-dissipative
and there is therefore no loss of information. Non-dissipative
interactions are expected when gravity is embedded in string
theory, so that an $S$-matrix description is possible. The
existence of an $S$-matrix makes it no longer automatic that, {\em
e.g.,} the $B$-violating  $\overline q l$ and $B$-conserving $qq$
outgoing channels have the same probability, as they would in
thermal evaporation. Thus, the problem of avoiding rapid baryon
decay or large Majorana neutrino masses is shifted to the
examination of symmetries~\cite{Krauss:1988zc} in the underlying
string theory which would suppress the appropriate
non-renormalizable operators at low energies.

I would like to thank Luis Epele, Jonathan Feng, Haim Goldberg, Concha
Gonzalez Garcia, Francis Halzen, Dan Hooper, Subir Sarkar, Al Shapere,
and Tom Weiler for valuable discussions.

\end{document}